\documentclass[amsmat,amssymb,amsfonts,aps,prb,twocolumn,showpacs]{revtex4}

\usepackage{graphicx}
\usepackage{dcolumn}
\usepackage{bm}
\begin{document}

\title{Magnetic and orbital blocking in Ni nanocontacts}

\author{ David Jacob, J. Fern\'andez-Rossier, J. J. Palacios }
\affiliation{Departamento de F\'{\i}sica Aplicada, Universidad de
Alicante, 03690 San Vicente del Raspeig, SPAIN }

\date{\today}

\begin{abstract}
We  address the fundamental question  of whether  magneto-resistance (MR) of
atomic-sized contacts of Nickel  is very large because of the formation of a
domain wall (DW) at the neck.  Using  {\em ab initio} 
transport calculations we find that, as in the case of
non-magnetic electrodes, transport in Ni nanocontacts depends very much on the
orbital nature of the electrons. Our results are in agreement with several
experiments in the average value of the conductance. On the other hand,
contrary to existing claims, 
DW scattering does {\em not} account for large MR in Ni nanocontacts.   
\end{abstract}

\maketitle


The strong sensitivity of the current flow  between two ferromagnetic metals
(FM's) separated by a non-magnetic region to the relative orientation of their
magnetization vectors is a fundammental physical phenomenon with a huge impact
in the magneto-electronics industry\cite{Wolf:science:01}. The figure of merit is the
ratio between the conductance for antiparallel (AP)  and parallel (P)
relative  magnetic orientations of the FM's, $x\equiv \frac{G_{\rm AP}}
{G_{\rm P}}$, 
which can be selected with an external magnetic field. Two different
conventions are used to characterize the so called magneto-resistance (MR),
MR$_1=100\times(1-x)\%$ and MR$_2=100\times(\frac{1}{x}-1)\%$.   Metallic
multilayers with alternated ferromagnetic and non-magnetic metallic layers
display a large MR known as Giant MR (GMR)\cite{Baibich:prl:88}.  MR in
FM/insulator/FM systems is known as Tunnel MR (TMR)\cite{Julliere:pl:75} and values
of MR$_{1}\simeq 30\%$ have been reported\cite{Moodera:prl:95}.  

More recently a number of groups
\cite{Garcia:prl:99,Chung:prl:02:Chopra:prb:02:Hua:prb:03,Sullivan:04,Oshima:apl:98:Ono:apl:99,Viret:prb:02,Gabureac:prb:04,Egelhoff:jap:04}
have studied MR in break junction systems, where  two sections of a Ni wire are
connected through an atomic-size contact. In this arrangement the intermediate
region connecting the  two bulk FM's  has a different geometry 
 but the same chemical composition, in contrast to GMR and TMR systems.  Some
groups have obtained  values of MR$_2$ going from  10$^2$ to 10$^4$  
($x\simeq  10^{-1}$ to $x\simeq 10^{-3}$)   
\cite{Garcia:prl:99,Chung:prl:02:Chopra:prb:02:Hua:prb:03,Sullivan:04} 
while others obtain moderate or even {\em negative}
values\cite{Oshima:apl:98:Ono:apl:99,Viret:prb:02,Gabureac:prb:04,Egelhoff:jap:04}.
In nanocontacts the resistance  
predominantly comes from the region with the smallest section, 
where electron transport is coherent and conductance is dominated by
the quantum mechanical transmission of electrons at the Fermi level\cite{Datta:book:95}.  
Mainly, two different mechanisms
have been proposed so far to account  for the  large values of MR, when
observed: Domain-wall (DW) scattering\cite{Tatara:prl:99,Imamura:prl:00} and
magnetostriction\cite{Gabureac:prb:04,Egelhoff:jap:04}. For the former it has been argued
that  in the AP arrangement a DW  is pinned at the atomic-size
contact\cite{Bruno:prl:99} and is responsible for strong spin scattering which gives
an extra contribution to the resistance as compared to the P configuration,
resulting  in a large ``ballistic'' MR\cite{Tatara:prl:99,Imamura:prl:00}. 

The fundamental question of whether MR is dramatically enhanced in atomic sized
ferromanetic contacts due to the presence of a DW remains open and is the
subject of this paper. Three ingredients are essential to answer this question.
First, as in the case of non-magnetic nano-contacts, the {\em electronic}
structure of the last atom(s) which determines the number of available
transport channels\cite{Agrait:pr:03}. Second,  the presence of
inhomogeneous magnetization profiles, e.g.,  a DW, which can induce
spin scattering and affect current
flow\cite{Tatara:prl:99,Imamura:prl:00,Levy:prl:97,Tatara:prl:97,vanHoof:prb:99}. Third, the {\em atomic
structure} (geometry)  which affects both the electronic and magnetic
structures, and thus, the transmission of these channels.  Previous theoretical
works present mutually  conflicting results with methodologies that either used
an oversimplified description of Ni electronic
structure\cite{Tatara:prl:97,Rocha:prb:04,Velev:prb:04} or idealized
geometries\cite{Smogunov:ss:02:Smogunov:ss:03,vanHoof:prb:99}.  Here we present transport calculations
across Ni nanocontacts describing the electronic, magnetic, and atomic
structure with ab-initio calculations\cite{Palacios:prb:01,Palacios:prb:02,Louis:prb:03}. Our results lead us to  conclude  that {\em intrinsic}
ballistic  MR is certainly {\em not} large in Ni nanocontacts. 

{\em Spin-dependent transport formalism.--}
Transport through atomic-size metallic contacts is currently understood  in
terms of elastic transport of non-interacting quasiparticles through a one-body
potential that describes  their interaction with the constriction. In this
approach, the conductance $G$ is proportional to the quantum mechanical
transmission ${\cal T}$ associated with the potential.  On the other hand,  the
spontaneous breaking of the spin degeneracy in  transition metal ferromagnets,
which is  due to electron-electron interactions, can also be properly
understood  in terms of a mean-field description, where quasiparticles interact
with a spin-dependent self-consistent potential. Once the self-consistent field
is determined for a given geometry, the quantum-mechanical spin-dependent and
energy-dependent transmission probability ${\cal T_{\sigma\sigma'}}(E)$  can be
obtained,  and thereby  the zero-bias conductance, using Landauer's
formula\cite{Datta:book:95}:
\begin{equation}
G=\frac{e^2}{h}[{\cal T_{\uparrow\uparrow}}(E_{\rm F})+{\cal T_{\downarrow\downarrow}}(E_{\rm F})+{\cal T_{\uparrow\downarrow}}(E_{\rm F})+{\cal T_{\downarrow\uparrow}}(E_{\rm F})].
\label{conductance}
\end{equation}
In the above expression we only make explicit the 
dependence of ${\cal T}_{\sigma\sigma'}$ on the spin channels, which we assume well-defined in the leads. 

{\em Ab initio Cluster Embedded Calculations.--}
It is an experimental
fact that the chemical nature of the contact determines the
conductance\cite{Agrait:pr:03}. As a rule of thumb, the
conductance of single-atom metallic nanocontacts can be as large as the number
of valence orbitals, but, in practice, is never larger than the number of
valence electrons. A natural description of the problem is in
terms of a localized atomic orbital basis, preferebly starting from first
principles\cite{Gaussian:03}. 
In previous publications\cite{Palacios:prb:01,Palacios:prb:02,Louis:prb:03}  we have presented
a method to perfom \textit{ab initio}  calculations  of quantum transport
through atomic constrictions and molecules which is based on the  code
GAUSSIAN\cite{Gaussian:03}. Our approach  has been successful in
explaining experimental results in paramagnetic
nanocontacts\cite{Palacios:prb:02,Garcia:prb:04}.  Here we take it a step further to
study systems without spin degeneracy, like ferromagnetic nanocontacts. 
We solve the problem dividing the system in three different parts: left (L) and
right (R) electrodes on one side and the contact region on the other.  The
spin-dependent one-body Hamiltonian is assumed fixed and homogeneous in the
bulk electrodes, but it is determined self-consistently in the contact region
subject to the appropriate magnetic boundary conditions.  

The density functional theory (DFT) calculations for the contact region are done 
with both LSDA and the hybrid functional B3LYP\cite{Gaussian:03}. 
The LSDA results are robust against different basis sets
and so we rely here on a minimal basis set with a core 
pseudopotential as described in previous work\cite{Palacios:prb:01,Palacios:prb:02}.
On the other hand, the B3LYP functional is more sensitive to the basis set
due to its non-local exchange contribution.
Therefore we employ here an all-electron basis set\cite{Doll:ss:03}.
The electrodes are described by means of a semi-empirical tight-binding Bethe
lattice model. With the appropriate parameters, the Bethe lattice can provide a geometry-independent
description of the contacts with a bulk density of states (DOS)  which is
smoother than the real DOS and mimics an average over both  disorder 
realizations and  the actual electrode crystal orientations.  
Spin-mixing solutions are not considered, i.e., $S_z$ is a good quantum
number. Thus, the last two terms in Eq.\ \ref{conductance} do not give any
contribution to the conductance. DW-like configurations are obtained for the
adequate magnetic boundary conditions and the constraint $S_z=0$.

\begin{figure}
\includegraphics[scale=0.3]{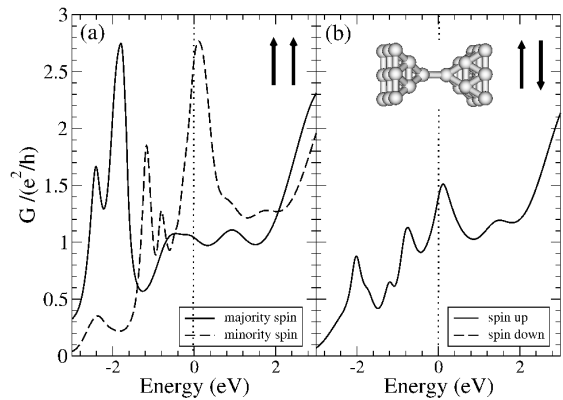}
\caption{(a) LSDA conductance per spin channel in the P configuration for the 
model nanocontact shown in the inset. (b) Same as in (a), but for the 
AP configuration.}
\label{symm-lsda}
\end{figure}

\begin{figure}
\includegraphics[scale=0.3]{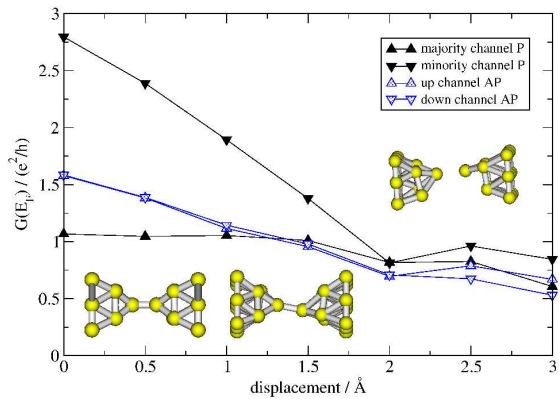} 
\caption{Conductance of both spin channels for the P and 
  AP configuration as a function of the stretching.
  The lines connecting the points are just to guide the eye.
  The insets show the relaxed contact geometry at different
  values of the displacement 
  (0\r{A}, 1.5\r{A}, and 3\r{A} from left to right).}
\label{stretch}
\end{figure}

{\em Results.--} We restrict ourselves to the study of the last (first) 
plateau of conductance upon  stretching (electrodeposition)  as, e.g.,
in the experiment of Viret et
al.\cite{Viret:prb:02} (Sullivan et al.\cite{Sullivan:04}). 
A reference atomic structure of the contact region has been
initially  taken like that shown in the inset of Fig. \ref{symm-lsda}. Following
Viret {\em et al.}\cite{Viret:prb:02}, we consider the  narrowest (and
most important) region to consist of two pyramids facing each other,
formed along the (001) direction, and with the
two tip Ni atoms 2.6 \AA \, apart forming a dimer. Bulk atomic distances and
perfect crystalline order are assumed otherwise. {\em Ab initio} simulations of the breaking
process as the one shown in Fig. \ref{stretch} 
support this choice. We stress that the section of the nanocontacts varies 
in the direction of the current flow. 
This is the situation in real nanocontacts and differs from perfect 1-dimensional
systems, studied in Refs. \onlinecite{Smogunov:ss:02:Smogunov:ss:03}, and from bulk systems
studied by Van Hoof {\em  et al.}\cite{vanHoof:prb:99}. 
In this regard, the geometries proposed by Bagrets {\em et al.}\cite{Bagrets:prb:04} 
are closer to real nanocontacts, but are not backed up by experiments or simulations.  

From the LSDA DOS projected on the tip
atoms (not shown) we see that the $sp$ orbitals are spin splitted by less than 1 eV and that
the minority (m) electrons are hybridized with the $d$ levels, which are
present at the Fermi energy. As a consequence, the LSDA DOS for the  majority (M)
electrons at the Fermi energy  is significantly {\em smaller} than that for the m ones.
These results are compatible with  LSDA first-principles
calculations for systems with translational invariance 
\cite{Smogunov:ss:02:Smogunov:ss:03},  have been properly
accounted for 
in Anderson-like model Hamiltonians\cite{Rocha:prb:04}, but  are in
marked  contrast to  the $J_{sd}$  model, usually invoked to understand large
values of the MR in nanocontacts\cite{Tatara:prl:99,Imamura:prl:00}.
In these models the transmission of itinerant $s$
electrons is perfect in the ferromagnetic case while the $d$ electrons do not contribute
to the current since they are localized. The MR depends thus dramatically on the
ratio between the spin splitting of the conducting $s$ electrons
and the Fermi energy when a DW is present. A large spin splitting
needed to give large MR is, however, at odds with the actual Ni
band structure and this model must be ruled out from the outset 
to account for large MR in Ni nanocontacts.

\begin{figure}
\includegraphics[scale=0.3]{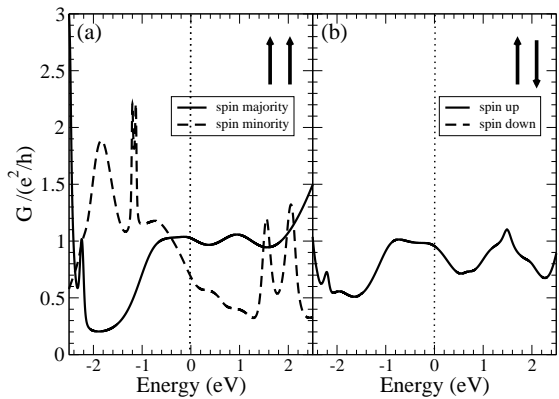}
\caption{(a) B3LYP conductance per spin channel in the P configuration for the 
model nanocontact shown in the inset of Fig. \ref{symm-lsda}. (b) Same as in (a), but for the 
AP configuration.}
\label{symm-b3lyp}
\end{figure}

Figure (\ref{symm-lsda}) shows the LSDA conductance as a function of energy for both up
and down spin channels in two situations: (a) Parallel (P) and (b) antiparallel (AP) bulk magnetic
arrangements.  In both cases the self-consistent solution has been forced to
respect the high symmetry of the  nanocontact. In the AP case the
self-consistent magnetization reverses abruptly between tip atoms. The resulting
magnetic moment for the contact atoms is $\approx 1.0 \mu_{\rm B}$ in both  
situations. This value is significantly larger than that obtained for bulk or surface atoms
($\approx 0.6 \mu_{\rm B}$
and reflects the low coordination of the tip atoms forming the contact. 
 In the  P
case the M channel is, for the most part, composed of a single $sp$ orbital channel and conducts
perfectly around the  Fermi energy (set to zero) while the m channel is composed of three orbital channels 
(one $sp$- and two $d$-like, which conduct roughly the same), 
and exhibits a transmission strongly dependent on the scattering energy. In the AP case the
system is invariant under the combined transformations that exchange
 L with R and 
$\uparrow$ with $\downarrow$, resulting in identical  values for the conductance of the two spin
channels, which now are composed of a dominant $sp$ channel and a strongly diminished contribution 
of the $d$ channels.
The conductance ratio for this particular case is $x=2.8/3.65=0.77$. This 
yields MR$_1=23\%$ and MR$_2=30\%$, which is clearly below large MR claims\cite{Garcia:prl:99,Sullivan:04}.

LSDA provides  a commonly accepted description  of the
electronic structure of bulk and surface ferromagnetism  in transition metals 
\cite{Doll:ss:03}.  However, the low coordination of atoms in nanocontacts
might give rise to a further localization of the $d$ electrons (compared with
bulk)  and an increase of the magnetic moment. Since LSDA fails to describe
properly localized  electrons due to the self-interaction problem, various alternatives
have been proposed to overcome this problem\cite{Wierzbowska:04}, the most popular being 
LDA+U\cite{Anisimov:prb:91} and SIC\cite{Perdew:prb:81}. 
We should point out that, while the results for the conductance reflect
the DOS and look plausible,
the fact that reported conductance histograms never show the lowest peak around 4$e^2/h$ at high
magnetic fields\cite{Untiedt:prb:04} make us
suspect that either the chosen 
model for the atomic structure is not realistic or that the electronic structure given by
LSDA, due to the problems mentioned above, does not provide the best approximation for nanocontacts. We explore below both possibilities separately. 

An alternative approach to the electronic structure comes from the use of a hybrid
functional like B3LYP which is a combination of Hartree-Fock and LSDA. The
former is free from the self-interaction problem, but fails to include correlation,
which is provided by the latter.  B3LYP happens to give a very good description
of the electronic structure and local magnetic moments in NiO\cite{Moreira:prb:02} and
La$_2$CuO$_4$\cite{Perry:prb:01}. 
With B3LYP the results for the conductance  (see Fig. \ref{symm-b3lyp}) 
are remarkably different in regard to the m channel. 
Now the $d$ channels give a much smaller contribution to the transmission at the Fermi energy. 
In this case the MR is negative and its absolute value is even smaller (MR$_1=-11\%$) 
than the one obtained with LSDA. 
With B3LYP the bulk and surface magnetic moments are slightly higher 
than the LSDA ones while the magnetic moment for the tip atoms is 
roughly the same ($\approx 1 \mu_{\rm B}$).

Since the m conductance evaluated 
at the LSDA level exhibits a strong dependence on the scattering energy,
we study now whether or not different
geometries  can change the  above results qualitatively. In an attempt to  explore
other realizations of the self-consistent potential compatible with the
magnetic boundary conditions and the experimental information, we  perform {\em
ab initio} structural relaxations as a function of the displacement between
outer planes in the core cluster. To do so,
 we consider a cluster like that shown in Fig. \ref{symm-lsda}.   The inner atoms in
the cluster (10 in total) are allowed to relax to local minimum energy
configurations as we stretch. This results, logically, in lower energy solutions and in the
loss of symmetry, so that the transmission in the AP case now becomes 
slightly spin-dependent.
In Fig. \ref{stretch} the conductance at the Fermi energy per spin
channel for the P and the AP configurations are shown  as a function of
the stretching up to the break-up point, starting  from a slightly compressed
nanocontact. From this figure
we see that the conductance of the m channel for the P configuration
changes significantly upon small changes. 
The MR, on the contrary,
barely changes as the nanocontact is stretched and is small, reaching vanishing 
values for the last points in Fig. \ref{stretch}. The conductance approaches a stable value
around $2e^2/h$ for both P and AP configurations.

{\em Discussion and conclusions.- } As mentioned earlier, the maximum number of
conducting channels in atomic-size contacts is roughly determined by the number
of valence electrons of the contact atom(s). However, as shown above, this
hypothetical upper limit is never reached, particularly for the m electrons,
remaining essentially only one M channel and {\em one} m channel transmitting
in the P case for stretched contacts.  
This result is impossible to predict without a full atomistic
self-consistent calculation. The M channel is $sp$-type.  Thus, this channel
transmits almost perfectly and evolves
smoothly with the stretching of the contact giving a stable contribution 
$ \cal{T}_{\uparrow\uparrow} \approx $ 1 (see Fig. \ref{stretch}. 
The $sp$ orbitals in the m channel
are  strongly hybridized with
$d$-orbitals and, therefore, are more sensitive to the contact geometry. 
The contribution to the conductance of the latter, which form narrower bands, 
disappears with the stretching and disorder, as expected (see Fig. \ref{stretch}).
On the other hand, in the AP configuration mostly one $sp$ orbital channel per spin
contributes.  The conductance {\em per spin channel} lies thus in the vicinity of $
e^2/h$, giving $\approx 2 e^2/h$ in total and is fairly stable during the
last stage of the breaking of the nanocontact. 

To conclude, the reason behind the very small MR values is the orbital (or geometric)
blocking of most of the \textit{a priori} available m channels in the P configuration due to 
the non-ideal geometry of the nanocontacts. The number of bands at the Fermi energy in the case
of a perfect mono-strand infinite ferromagnetic chain\cite{Smogunov:ss:02:Smogunov:ss:03} 
is much larger than the number of non-zero eigenvalues of the transmission matrix in a nanocontact 
(for a given basis and functional). This phenomenon affects mainly the $d$ bands (no DW involved) and, 
therefore, we call it orbital blocking. On the other hand, $sp$ bands are less sensitive to geometry.
In fact, there can only be a significant and positive MR  in the cases where the m
channel in the P configuration conducts appreciably. This is  the case for the highly 
symmetric nanocontact presented in Fig. \ref{symm-lsda} within LSDA and, e.g., for the Ni chains 
studied in Refs. \onlinecite{Smogunov:ss:02:Smogunov:ss:03}. In the P case, the number of M and m bands at
the Fermi energy is 1 and 6, respectively. When a DW is formed
in the chain, 5 out of 6 m channels are blocked, remaining 1 per spin, almost
fully transmitting. As a result, one expects a ratio $x\simeq2/7$ resulting in
MR$_{2}\simeq250\%$ for the ideal chain\cite{Smogunov:ss:02:Smogunov:ss:03}. 
The blocking of a number of channels
is due to the fact that the $d$ electrons, and not the $sp$ electrons, are spin
splitted. This could explain some early results\cite{Garcia:prl:99}, but not recent ones\cite{Sullivan:04}. 
Furthermore, to date, no evidence of chain formation in Ni has been
reported. Even so, scattering at the electrode-chain contact will always be
present. 
For completeness, we have also performed calculations for Ni chains using the B3LYP functional. 
The number of bands crossing the Fermi energy is now reduced to 1+4 compared to the LSDA results.
This agrees with recent LSDA+U calculations reported by Wierzbowska et al.\cite{Wierzbowska:04}
where the two degenerate flat minority bands $d_{xy},d_{x^2-y^2}$ are shifted downwards in energy
because of the exchange interaction cancelling part of the self-interaction of the 
strongly localized electrons in these flat bands.
However, this is not the reason for the drop in the m conductance seen in Fig.\ref{symm-b3lyp}
since the corresponding channel does not contribute to the conductance in the LSDA case either.
In addition, non-collinear DW's, not considered here,
also reduce the MR\cite{Jacob:05}. It is our believe that
when observed, large values of the MR in Ni nanocontact might be due to
magnetostriction effects and the  corresponding formation of wider section
contacts or to the presence of adsorbates, which modify the local electronic
structure\cite{Untiedt:prb:04}.


We acknowledge E. Louis, C. Untiedt, J. A. Verg\'es, and
G. Chiappe for fruitful discussions. 
JJP acknowledges financial support from 
Grants No. 1FD97-1358 (FEDER funds) and No. MAT2002-04429-C03 (MCyT).
DJ  acknowledges financial support from MECD under grant UAC-2004-0052.
JFR acknowledges financial support from Grant No.
MAT2003-08109-C02-01 (MCyT), No. UA/GRE03-15 (Universidad de Alicante), and
Ramon y Cajal Program (MCyT).


\bibliography{matcon}

\end{document}